\documentclass[twocolumn]{aastex631}

\usepackage{amsmath}
\usepackage{bm}
\usepackage[]{threeparttable}

\received{February 23, 2022}
\accepted{October 29, 2022}

\shorttitle{Zonal winds of Uranus \& Neptune}
\shortauthors{Soyuer et al.}
\graphicspath{{./}{figures/}}

\begin{document}

\title{Zonal winds of Uranus and Neptune:\\ Gravitational harmonics, dynamic self-gravity, shape, and rotation}

\author[0000-0002-8238-9747]{Deniz Soyuer}

\author[0000-0002-5794-0453]{Benno Neuenschwander}

\author[0000-0001-5555-2652]{Ravit Helled}

\affiliation{Center for Theoretical Astrophysics and Cosmology,
Institute for Computational Science, \\
University of Zurich,
Winterthurerstrasse 190, 8057 Zurich, Switzerland}

\email{deniz.soyuer@uzh.ch}



\begin{abstract}
Uranus and Neptune exhibit fast surface zonal winds that can reach up to few hundred meters per second. Previous studies on zonal gravitational harmonics and Ohmic dissipation constraints suggest that the wind speeds diminish rapidly in relatively shallow depths within the planets.  Through a case-by-case comparison between the missing dynamical gravitational harmonic $J^\prime_4$ from structure models, and with that expected from fluid perturbations, we put constraints on  zonal wind decay in Uranus and Neptune.
To this end, we generate  polytropic empirical structure models of Uranus and Neptune using $4^{\rm th}$--order Theory of Figures (ToF) that leave hydrostatic $J_4$ as an open parameter. Allotting the missing dynamical contribution to  density perturbations caused by zonal winds (and their dynamic self-gravity), we find that the maximum scale height of zonal winds are $\sim 2-3\%$ of the planetary radii for both planets.  Allowing the models to have $J_2$ solutions in the $\pm 5 \times 10^{-6}$ range around the observed value has similar implications.
The effect of self-gravity on $J^\prime_4$ is roughly a factor of ten lower than that of zonal winds, as expected. The  decay scale heights  are virtually insensitive to the proposed modifications to the bulk rotation periods of Uranus and Neptune in the literature.
Additionally, we find that the dynamical density perturbations due to zonal winds have a measurable impact on the shape of the planet, and could potentially be used to infer wind decay and bulk rotation period via future observations. 
\end{abstract}



\section{Introduction} \label{sec:intro}
Uranus and Neptune are the least explored giant planets in the solar system and have been only visited once by  \textit{Voyager II} in the late 1980s. Due to the lack of \textit{in situ} measurements, there are ambiguities in various features in the two planets such as; their composition \citep{nettel, helled_comp}, their heat transfer \citep{podolak, vazan, bailey, 50megabyte_paper}, presence of boundary layers \citep{helled, nettel}, their geopotential models \citep{ j_nep, j_ura}, origin of their peculiar magnetic fields \citep{holme, stanley1, stanley2, krista1, krista2}, their zonal wind structure and its decay profile \citep{kaspi2013, soyuer2020, gcm} and the zonal wind--magnetic field back--reaction \citep{soyuer2021}.

Predicting the structure and the dynamics of atmospheric circulation in Uranus and Neptune (both zonal and meridional) is crucial for understanding their heat transfer mechanisms \citep{podolak, vazan, 50megabyte_paper} and in turn their composition and vertical mixing of molecules \citep{nettel, bailey, atmos}, generation of secondary magnetic fields via their interaction with zonal winds \citep{hao, soyuer2021}, and molecular advection phenomena \citep[e.g. like those recently shown in Jupiter with ammonia][]{ferrel}.

\begin{figure}
    \centering
    \includegraphics[width = \columnwidth]{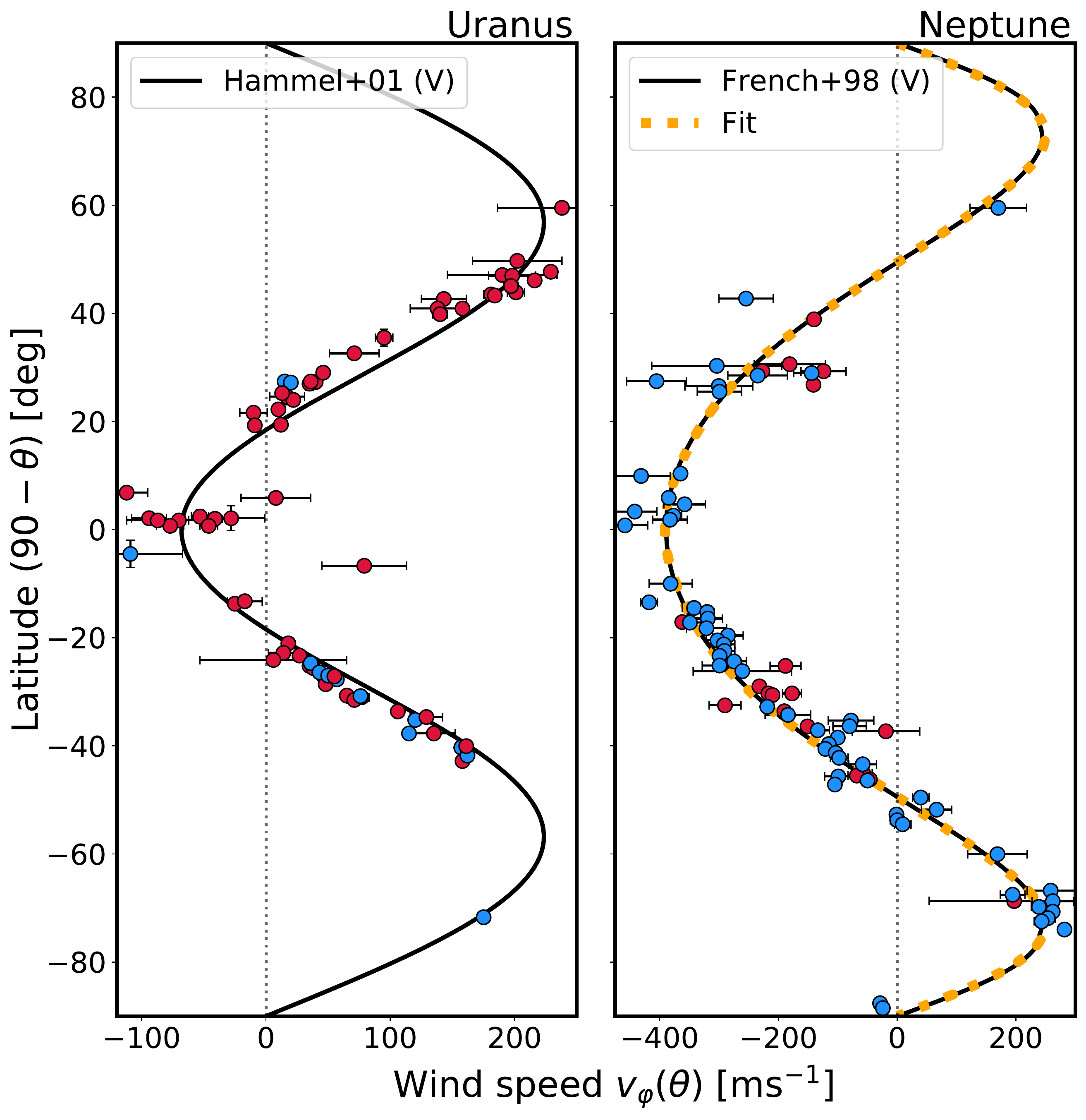}
    \caption{\textbf{Surface zonal wind speeds of Uranus and Neptune as a function of latitude.} On the left panel; red points show Keck and Hubble Space telescope measurements \citep{hammel_keck, sromovsky} and blue points show Voyager II measurements \citep{hammel}. The solid line is the fit by the latter. On the right panel; red points show Hubble Space telescope measurements \citep{sro} and blue points show Voyager II measurements \citep{lindal, limaye}. The solid line is the fit by \citet{french}, and the orange dotted line is our fit given by equation ($\ref{eq:sin}$).}
    \label{fig:winds}
\end{figure}

Uranus and Neptune are both observed to have fast surface zonal winds with speeds reaching up to  200 ms$^{-1}$ and 400 ms$^{-1}$ relative to their respective assumed bulk rotations, as shown in Figure \ref{fig:winds}. How exactly the winds decay with depth is uncertain, 
but there is evidence  that they must decay rapidly in the shallow layers of Uranus and Neptune \citep{kaspi2013, soyuer2020}. 
In the pioneering study of  \citet{kaspi2013} the depths of the winds were constrained by comparing the effects of dynamical density perturbations due to winds to existing zonal harmonics data. In this paper, we perform a follow up study
by; \textit{(i)} using polytropic structure models generated with 4$^{\rm th}$-order Theory of Figures (ToF) (instead of polynomial models generated by 3$^{\rm rd}$-order ToF), \textit{(ii)}  using up-to-date zonal gravitational harmonics data \citep{j_ura}, \textit{(iii)} doing a case by case analysis for each model's decay height, \textit{(iv)} including the effects of self-gravity (discussed in \citet{galanti2017, selfg}), \textit{(v)} probing the effects of rotation period variations from \textit{Voyager II} measurements and \textit{(vi)} interpreting zonal wind-induced dynamical density perturbations in the context of planetary shapes.

Our paper is structured as follows: In Section \ref{sec:methods} we present the methods for calculating the gravitational harmonic perturbations and the density perturbations associated with them, along with a short summary of self-gravity. Section \ref{sec:interior} elaborates on our empirical structure models and their generation procedure. Then, in Section \ref{sec:res} we present our results on decay scale heights and the influence of dynamic density perturbations on planetary shapes. Lastly, we discuss our findings and finish with our concluding remarks in Sections \ref{sec:disc}.

\section{Methods}
\label{sec:methods}

\subsection{Zonal gravity harmonics \texorpdfstring{$J_n$}{Lg}}
The effective external potential of a uniformly rotating planet is given by the contribution of its gravitational field, and its centrifugal potential due to bulk rotation:
\begin{align}
&\quad\quad\quad\quad\quad\quad\Psi_e(r,\theta) = \Psi_g (r,\theta) + \Psi_\omega (r,\theta)  \\ \nonumber
&= -\frac{GM}{r} \left(1 - \sum\limits_{n=1} \left(\frac{R}{r} \right)^n J_{n} P_{n}(\cos\theta)\right) - \frac{1}{2}\omega^2r^2 \sin^2\theta,
\end{align}
where $M$ is the planet's mass, $G$ the gravitational constant, $r$ the radial parameter, $R$ the planetary radius (taken as the 1-bar radius in this work), $\omega$ the assumed bulk rotational velocity, $J_n$ are the zonal gravity harmonics of the geopotential model, and $P_{n}$ are the Legendre polynomials.
The first term on the right hand side describes the gravitational potential, in which $-GM/r$ is the gravitational potential of a perfect sphere and the series expansion reflects the deviations that are symmetric with respect to the rotation axis (i.e., zonal deviations), where $J_n$ are the zonal gravitational harmonics, defined as:
\begin{equation}
J_n = -\frac{1}{M R^n}\int\limits_V  \rho(\mathbf{r}) P_n(\cos{\theta})  r^n \, {\rm d} V,
\label{eq:j_n}
\end{equation}
and the integral is over the entire  volume of the planet.
This picture is of course only valid if the whole planet is described by bulk rotation, which is not exactly the case for the solar system giants due to strong zonal winds, implying differential rotation down to some unknown depth. Thus, we can redefine the density distribution by separating it into a hydrostatic density $\tilde{\rho}$ (present in the absence of winds), and a perturbative term $\rho^\prime$ due to the existence of winds, such that the total density is the sum of both \citep{kaspi1}: 
\begin{equation}
    \rho = \tilde{\rho} + \rho^\prime.
\end{equation}
The added density perturbation $\rho^\prime$ implies a perturbation in the gravitational harmonics as well. We can say that the observed harmonics $J^{\mathrm{\scriptscriptstyle{obs}}}_n$ are the sum of the hydrostatic $\tilde{J}_n$ and the dynamic terms $J_n^\prime$ \citep{kaspi1}:
\begin{equation}
    J^{\mathrm{\scriptscriptstyle{obs}}}_n = \tilde{J}_n + J_n^\prime,
    \label{eq:presc}
\end{equation}
where $J^\prime_n$ is simply obtained by plugging the density perturbation $\rho^\prime$ into equation (\ref{eq:j_n}). The derivation of $\rho^\prime$ follows from perturbing the fluid momentum equation, described in detail below. For Uranus and Neptune only the even harmonics $J_2$ and $J_4$ have been determined \citep{j_nep, j_ura}. In this study we mainly focus on the perturbation to $J_4$. The case where we investigate the effect of varying $J_2$ on the zonal wind decay heights is also shown for completeness. We denote the gravitational harmonic perturbation $J^\prime_n$ calculated via including a density perturbation $\rho^\prime$ in the fluid  equations as: $J^\prime_n(\rho^\prime)$. 

We generate interior structure profiles of Uranus and Neptune that fit the measured $J^{\rm obs}_2$ zonal harmonic, and then take this as the hydrostatic $\tilde{J}_2$, while keeping $J_4$ as a free parameter. Since generated density profiles are by definition in hydrostatic equilibrium, one obtains $\tilde{J}_4$ as an output of this prescription. We can therefore  attribute the difference between the model output $\tilde{J}_4$ and the observed $J^{\rm obs}_4$ to the missing zonal wind contribution, which we denote as $J^\prime_4(\tilde{\rho})$ (i.e., the dynamical contribution that is not included in the hydrostatic model, similar to \citet{kaspi2013}):
\begin{equation}
J^\prime_4(\tilde{\rho}) = J^{\mathrm{\scriptscriptstyle{obs}}}_4 - \tilde{J}_4.
\end{equation}
This way, we can demand that both ways of calculating the perturbation should yield the same results:
\begin{equation}
J^\prime_4(\rho^\prime) \overset{!}{=} 
J^\prime_4(\tilde{\rho}),
\label{eq:primee}
\end{equation}
and then selectively model the phenomena that  contribute to the density perturbations in the momentum equation, explained below. Hence, different interior structure models would yield different dynamical contributions associated with various terms in the fluid momentum equation.

\subsection{Perturbing the momentum equation}
The Navier-Stokes momentum equation in a frame co-rotating with the planetary bulk rotation with added zonal wind contribution $\bm{U}_\varphi = U_\varphi \hat{\bm{\varphi}}$ can be written as \citep{selfg}:
\begin{equation}
\rho \frac{D\bm{U}_\varphi}{Dt} + 2 \omega \rho \hat{\bm{z}} \times \bm{U}_\varphi = -\nabla P - \rho \nabla \Psi_e +  \bm{j} \times \bm{B} + \nu \nabla\cdot \bm{S},
    \label{eq:ns}
\end{equation}
where $\rho$ is the density, $P$ the pressure, $\bm{j}$ the electrical current density, $\bm{B}$ the magnetic field, $\nu$ the viscosity, $\bm{S}$ the traceless rate-of-strain tensor, and $\Psi_e$ is the effective gravitational potential given by
\begin{equation}
    -\nabla \Psi_e = -\nabla (\Psi_g + \Psi_\omega) = -g \hat{\bm{r}} + r \omega^2 \sin\theta \hat{\bm{\rho}},
\end{equation}
where $\hat{\bm{\rho}}$ is the cylindrical radial unit vector.

In hydrostatic equilibrium, the effective potential $\tilde{\Psi}_e$ obeys hydrostatic balance and is a solution to the Poisson equation:
\begin{equation}
    \nabla\tilde{P} = -\tilde{\rho}\nabla\tilde{\Psi}_e, \,\,\,\,\,\,\,\,\, \nabla^2 \tilde{\Psi}_e = (4\pi G \tilde{\rho} - \omega^2).
\end{equation}
Expanding the buoyancy parts in the hydrostatic equilibrium with the perturbative terms and neglecting the second order perturbations, we get the relation
\begin{equation}
    -\nabla P - \rho\nabla\Psi_e = - \nabla P^\prime - \nabla(\tilde{\rho}\Psi^\prime_e) + \Psi^\prime_e\nabla\tilde{\rho} - \rho^\prime \nabla \tilde{\Psi}_e,
\end{equation}
where we have applied the chain rule to the term with the self-gravity $\Psi^\prime_e$.
In a steady state with $\partial_t U_\varphi = 0$, neglecting the contribution from the viscous and the magnetic terms, the  curl of equation ($\ref{eq:ns}$) reads
\begin{align}
    &\quad\quad\quad\tilde{\rho}\frac{2U_\varphi}{r}\left(\frac{\partial_\theta U_\varphi}{r} - \cot\theta \partial_r U_\varphi \right) -2 \omega \partial_z(\tilde{\rho} U_\varphi)=  \nonumber \\
    &
    \!\!\left\{ \nabla \times (\Psi^\prime_e \nabla \tilde{\rho}  - \rho^\prime \nabla \tilde{\Psi}_e ) \right\} \cdot \hat{\bm{\varphi}} = -\frac{1}{r}\partial_\theta (\Psi^\prime_g \partial_r \tilde{\rho}  - \rho^\prime \partial_r \tilde{\Psi}_g )
    \label{eq:curl},
\end{align}
in the linearized framework, where in the last step  we neglect the contribution by the centrifugal potential, such that ${\Psi}_e \approx {\Psi}_g$ for both the hydrostatic and the perturbative terms, which is not  strictly justified \citep{selfg}. 
Uranus and Neptune have a flattening of roughly $\sim \!\!2\%$ and therefore assuming spherical symmetry for the gravitational potential has a rather small  contribution to the overall result. As a result, the latitudinal derivatives of the hydrostatic terms arising in the curl of equation (\ref{eq:curl}) are also neglected. 
We note that  keeping only the self-gravity term in the equation is not a full prescription and that the effect of  additional terms  is of the same  order of magnitude.  We include the self-gravity term, which is  easier  to include, in order to demonstrate the maximum contribution it has on the momentum equation.

The density perturbations associated with Lorentz forces are calculated using the electrical conductivity prescription given in \citet{soyuer2020} and become comparable in strength to those of the zonal winds, however, only within far deeper regions. The contribution of Lorentz forces  to  $J_4^\prime$ through  magnetic density perturbations is  found to be at least 4-5 orders of magnitude lower than that of zonal winds, and therefore negligible.
Re-writing $\partial_r \tilde{\Psi}_g = \tilde{g}$, and integrating over the co-latitude and rearranging the terms, we can isolate the density perturbation:
\begin{equation}
    \rho^\prime =  \frac{1}{\tilde{g}}\left(r \int \mathcal{U}(U_\varphi, \tilde{\rho}) \, {\rm d}\theta - \Psi_g^\prime \partial_r\tilde{\rho} \right),
    \label{eq:rho_prime}
\end{equation}
where the zonal wind terms $\mathcal{U}(U_\varphi, \tilde{\rho})$ (denoting  the left hand side of equation (\ref{eq:curl})) and self-gravity terms are separated. Note that the integral over $\theta$ also has an integration constant $\rho_0(r)$ that only depends on $r$. In spherical symmetry, this integration constant has no effect on the zonal gravitational harmonics calculated via equation (\ref{eq:j_n}).

The Poisson equation for self-gravity  $\Psi^\prime_g$  can be reformulated as an inhomogeneous Helmholtz equation: 
\begin{equation}
    \bigg(\nabla^2  + \underbrace{4 \pi G  \frac{\partial_r\tilde{\rho}}{\tilde{g}}}_{:= \mu(r)}\bigg)\Psi_g^\prime = 4\pi G \bigg(\underbrace{\frac{r}{\tilde{g}}\int \mathcal{U}(U_\varphi, \tilde{\rho}) \,  {\rm d}\theta}_{:=\rho^\prime_U}\bigg),
    \label{eq:helmholtz}
\end{equation}
which is (barring the inertial terms) equivalent to the formulation in \citet{selfg}, where the terms under the braces are similar to the notation in that work.

As mentioned in \citet{selfg}, assuming a constant $\mu$ considerably simplifies the problem. Making this approximation and solving for the self-gravity potential $\Psi_g^\prime$, which we can plug back into equation (\ref{eq:rho_prime}), we get the total density perturbation $\rho^\prime = \rho^\prime_U + \rho^\prime_{\Psi_g^\prime}$, and subsequently $J_4^\prime(\rho^\prime)$. For each density profile, we can repeat this scheme for different zonal wind decay parameters until $J_4^\prime(\rho^\prime)$ matches $J_4^\prime(\tilde{\rho})$.

\subsection{Calculating the effect of dynamic self-gravity}
We derive the effect of the self-gravity potential $\Psi^\prime_g$ following \citet{selfg} as shown in Appendix \ref{sec:app}. Here, we briefly elaborate on the results, and refer the reader to their work for a more detailed picture. Hereafter, we omit the subscript of the self-gravity potential.

\subsubsection{Estimating \texorpdfstring{$\mu$}{Lg}}
\begin{figure*}
    \centering
    \includegraphics[width = \textwidth]{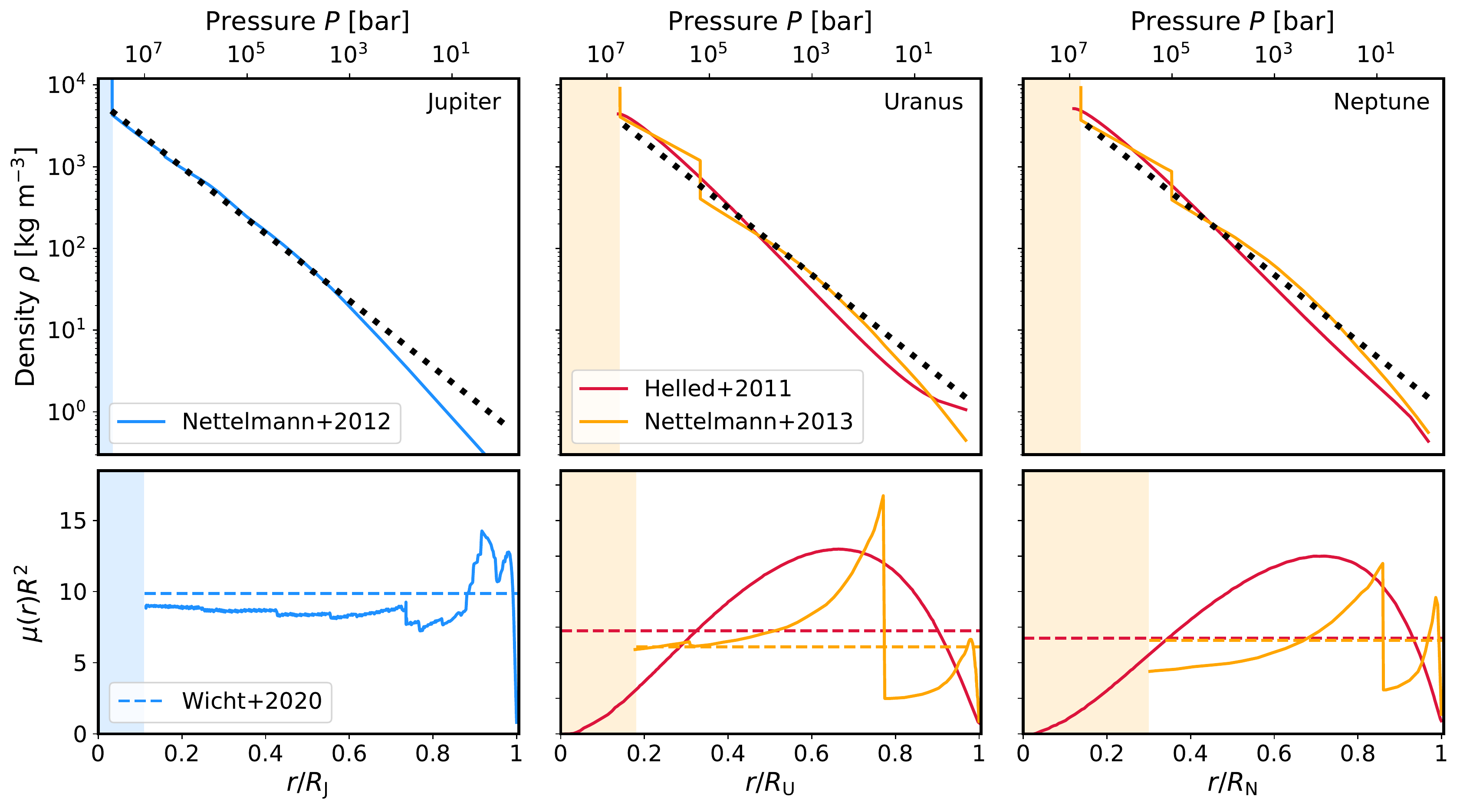}
    \caption{\textbf{Top panels: Pressure vs. density profiles of various published interior structure models}. Solid lines show the pressure vs. density profiles of the models, where the highlighted regions show the cores of the models, if it exists. 
    The dotted black lines are arbitrarily normalized polytropes with a polytropic index of $n=1$ to guide the eye.
    For Jupiter, a polytropic index $n$ of unity is a relatively good fit for the model of \citet{nettel_jup} beneath 1 kbar ($\sim 0.99R_{\rm J}$), as also shown in \citet{selfg}. For Uranus and Neptune  $n = 1$ seems to be a somewhat decent fit \cite{helled,nettel}. 
    \textbf{Bottom panels:} Solid lines show the calculated $\mu(r)R^2$ for all models. The dashed line in the bottom left panel shows the $\mu \sim \pi^2/R_{\rm J}^2$ approximation made by \citet{selfg} for Jupiter. The dashed lines in the other bottom panels show the average value of $\mu(r)R^2$ throughout the core exterior calculated here. Due to the relative complexity of the structure models shown here, the $\mu(r)R^2$ value has non-monotonous behaviour, which is not the case for the polytropic models we consider in the following chapters.}
    \label{fig:mu}
\end{figure*}

As shown in the Appendix \ref{sec:app} the self-gravity potential $\Psi^\prime$ can be expressed as a series of normalized spherical Bessel functions $j^\star_{ln}(r) = N_{ln} j(k_{ln}r)$:
\begin{equation}
\Psi^{\prime}(\bm{r})=\sum_{n=1}^{\infty} \sum_{l=1}^{\infty} \Psi_{l n}^{\prime} j_{l n}^{\star}(r) P_{l}(\cos\theta).
\end{equation}
The coefficients $\Psi^\prime_{ln}$ are then found via the spectral decomposition, acquired by plugging the potential into the inhomogeneous Helmholtz equation (\ref{eq:helmholtz}), and exploiting the orthonormality of $j^\star_{ln}(r)P_l(\cos\theta)$:
\begin{equation}
\Psi_{l n}^{\prime}=-\frac{G(2 l+1)}{k_{l n}^{2}-\mu} \int \rho^\prime_U(\bm{r}) j_{l n}^{\star}(r) P_{l}(\cos\theta) {\rm d}V.
\end{equation}
A fundamental assumption that this scheme makes is that the inhomogeneity in the Helmholtz equation $\mu$ is constant. Figure \ref{fig:mu} shows the pressure $P$ vs.  density $\rho$, and $\mu(r)$ as a function of normalized radius for the Jupiter model by \citet{nettel_jup} used in \citet{selfg}, as well as for two sets of different published structure models of Uranus and Neptune  \citep{nettel_jup,nettel,helled}. For the Jupiter model, as well as the Uranus and Neptune models, $\mu$ varies significantly in the outer regions.

\subsection{Zonal winds}
The surface profile of Uranian winds is described by: 
\begin{equation}
v_{\varphi,\mathrm{\scriptscriptstyle{U}}} (\theta) =  170 \times \left(0.6\sin\theta + \sin3\theta\right) \,\,\mathrm{m s^{-1}},
\label{eq:w_ura}
\end{equation}
with $\theta$ the co-latitude \citep{hammel}. For Neptune, the surface wind profiles expressed in degrees in latitude ($\phi = 90^\circ - \theta$) reads
\begin{equation}
v_{\varphi,\mathrm{\scriptscriptstyle{N}}} (\phi) = -389+0.188\phi^{2}-1.2 \times 10^{-5}\phi^{4},
\label{eq:w_nep}
\end{equation}
with a high-latitude-taper applied to latitudes $|\phi| > 75^\circ$, such that the wind profile vanishes at the poles  \citep{sro, french}. Since the tapered Neptunian winds are cumbersome to deal with analytically, we approximate the above expression for Neptune with a sum of sine functions for ease of computation:
\begin{equation}
    v_{\varphi,\mathrm{\scriptscriptstyle{N}}} (\theta) \simeq  \sum_{i=1}^{6} c_i \sin((2i-1) \theta).
    \label{eq:sin}
\end{equation}
As can be seen in Figure \ref{fig:winds}, this simple fit matches the tapered fit very accurately. The coefficients $c_i$ are given by: $c_1 = -219, ~c_2 =  240.02, ~c_3 = 93.58, ~c_4 = 32.62, ~c_5 = 12.98, ~c_6 = 5.82$.

Surface winds are predicted to penetrate  along cylinders parallel to the planet's rotation axis. 
The conclusions drawn in \citet{kaspi2013} are suggestive of relatively shallow winds in both Uranus and Neptune. 
As shown \citet{soyuer2020}, deep-seated strong winds would couple too much to the background magnetic field (due to rapidly increasing electrical conductivity with depth), and cause excessive amounts of Ohmic dissipation, violating the energy/entropy budgets of the planets.
Fast decaying winds are also very likely in Jupiter and Saturn, where constraints on zonal wind decay from gravity measurements by Juno and Cassini  and  Ohmic dissipation constraints are in general agreement \citep{liu2008, wicht,kaspi2018,kaspi2020, galanti2020}.

We define an azymuthally symmetric zonal wind profile $U_\varphi (r,\theta, n_i)$, with a radial decay profile $\mathcal{B}(r, n_i)$:
\begin{equation}
U_\varphi (r, \theta, n_i) = v_\varphi(\mathcal{A}(r,\theta)) \times \mathcal{B}(r, n_i).
\label{eq:u}
\end{equation}
analogous to that in \citet{kaspi2013} and same as in \citet{soyuer2021} where $n_i$ are any parameters describing the decay.
$\mathcal{A}(r,\theta)$ is a functional that links any point $(r,\theta)$ within the planet to the surface wind profile $v_\varphi(r = R_\mathrm{\scriptscriptstyle{U,N}}, \theta)$, given by equations (\ref{eq:w_ura}) and (\ref{eq:w_nep}, \ref{eq:sin}). For wind profiles that run along lines parallel to the rotation axis, $\mathcal{A}(r,\theta)$ is given by:
\begin{equation}
   \mathcal{A}(r,\theta) = \arcsin\left(\frac{r \sin\theta}{R_\mathrm{\scriptscriptstyle{U,N}}} \right). 
\end{equation}
We adopt an exponentially decaying zonal wind profile with a scale height $H$:
\begin{equation}
\mathcal{B}(r, H) = e^{\left({r}/{R} -  1\right)/H} .
\end{equation}
where $R$ is the planetary radius. More sophisticated models with additional open parameters have been proposed for giant planet zonal wind decay \citep[see e.g.][]{kaspi2018}. However, since the zonal gravitational harmonics are integrated quantities of the density perturbations, and thus of the zonal winds, the exact profile of the decay has a negligible effect on the outcome of the total perturbation, since the winds are thought to decay significantly in the outermost regions anyway.

\section{Interior structure models}
\label{sec:interior}
\begin{figure*}
    \centering
    \includegraphics[width = \textwidth]{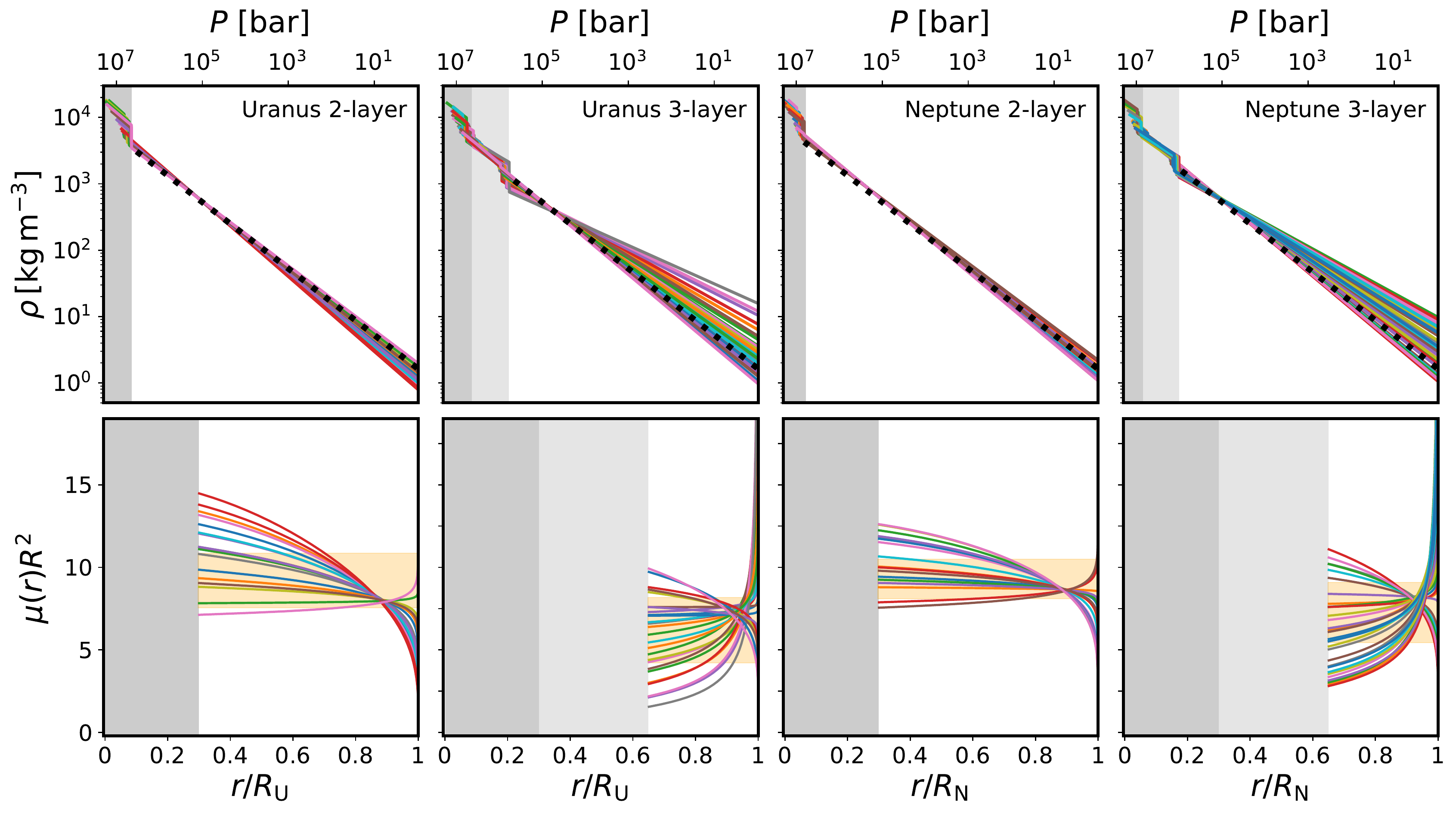}
    \caption{\textbf{Top panels: Pressure vs.~density profiles of Uranus and Neptune using polytropic empirical structure models where the innermost polytrope represents the core.}  The dotted black line represents a polytrope with index unity $P \propto \rho^2$ to guide the eye. \textbf{Bottom panels}: Solid lines represent the calculated $\mu(r) R^2$ values for each model and the shaded yellow region shows the range of mean values $\langle \mu(r)\rangle  R^2$  for all the empirical density models. The mean values are computed for the outermost polytrope only. Models with a steeper (flatter) $P$ vs.~$\rho$  curve than the index unity polytrope have a monotonously decreasing (increasing)  $\mu(r) R^2$ value.}
    \label{fig:poly}
\end{figure*}

We generate empirical structure models based on two or three piece-wise arranged polytropes following \citet{benno,benno2}. The pressure vs. density profiles of the models, along with their $\mu(r)R^2$ values is presented  in Figure \ref{fig:poly}. 

A polytrope relates the pressure $P$ and the density $\rho$ via the free polytropic index $n$ and the free polytropic constant $K$:
\begin{equation}
    P(\rho) = K \rho^{\frac{n+1}{n}}.
\end{equation}

We consider up to three regions represented by different polytrope allowing  different physical behavior. However, unlike in structure models that are based on physical equations of state, these regions do not necessarily exhibit constant and well-mixed composition. 
Through an iterative process a density profile is generated such that, if in hydrostatic equilibrium, obeys the polytropic relations. This density profile, together with the planet's mass, its equatorial radius and assumed uniform  rotation period is evaluated by an implementation of the $4^{\rm th}$--order Theory of Figures (ToF), developed and described in \cite{Zharkov1970,Zharkov1975,Zharkov1978,Hubbard2014,Nettelmann_2017}. Assuming hydrostatic equilibrium, this method evaluates its shape and even gravitational coefficients $J_2$, up to $J_8$. A cost function is introduced to control the free parameters $n_i$ and $K_i$ and the radii 
of transition between the polytropes, so the resulting model fits the planetary parameters as listed in Table \ref{tab:u_tab} \citep[for more details, see][]{benno}.

We assume that the observed zonal harmonic $J_2^{\rm obs}$ is equal to the hydrostatic $\tilde{J}_2$ used as a constraint in the models, and leave the hydrostatic model harmonic $\tilde{J}_4$ as an open parameter. This allows for more variation in the scale height parameter space, since $J_4$ is more sensitive to shallower regions than $J_2$, and also the measurement errors on $J_4^{\rm obs}$ are larger than $J_2^{\rm obs}$ for both planets.

We exclude models for Uranus and Neptune that show a central density  $\rho_{center} > 18'000 \text{ kg m}^{-3}$, which 
is well above the density of ``rock" at the expected core pressure and temperature \citep{Thompson1974, sesame7100, Musella2019}. For all the models, we set the maximum core radius as $0.3 R$, for both planets. However, the exact choice of the core boundary is insignificant when it comes to calculating the contribution of the mass in deeper layers, due to $J_4$'s insensitivity to deeper regions.
\begin{threeparttable}
	\centering
	\caption{Parameters used in generating empirical structure models of Uranus and Neptune via ToF. The $J_2$ values are renormalized according to their respective equatorial radii.}
	\begin{tabular}{l || c c } 
		\hline
		 Parameter & Uranus & Neptune \\
		\hline
		$J_2 \times 10^{-6}$ & 3510.68 $\pm$ 0.7\tnote{a} & 3535.94 $\pm$  4.5\tnote{b}  \\
		$M$ [$M_\oplus$] & 14.536\tnote{a} & 17.148\tnote{b} \\
		$T$ [h] & 17.24\tnote{c} & 16.11\tnote{d}  \\
		$R_{\rm eq}$[km] &  25559\tnote{c}  & 24766\tnote{d} \\
		\hline
	\end{tabular}
	\begin{tablenotes}
	    \item[a] \cite{j_ura}
	    \item[b] \cite{j_nep}
	    \item[c] \cite{Lindal1987}
	    \item[d] \cite{Lindal1992}
\end{tablenotes}
	\label{tab:u_tab}
\end{threeparttable}
Since $\sim\!\!90 \%$ of the  contribution to $J_4^\prime$ comes from the depths of $r = [0.65 R,R]$ within the planet \citep{benno}, we also generate 3-layer empirical density models where the inner boundary of the outermost polytrope starts from $<\!0.65 R$. The effect of the zonal wind and self-gravity induced $J_4$ perturbation in this region is therefore very similar to that by the entire planet, and allows us to  estimate the self-gravity contribution to $J^\prime_{4}$.

\begin{figure}
    \centering
    \includegraphics[width = \columnwidth]{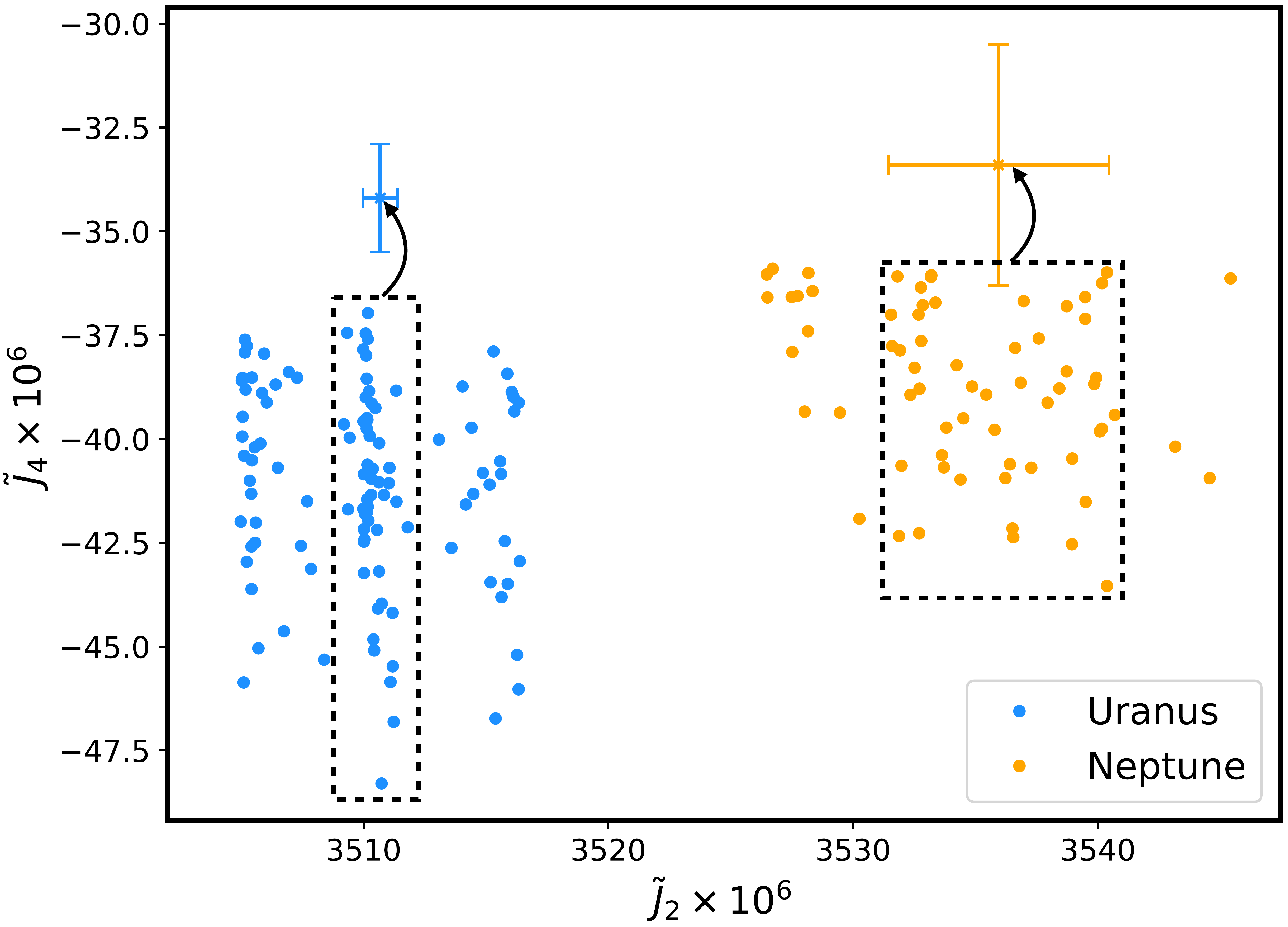}
    \caption{The $\tilde{J}_2$ and the $\tilde{J}_4$ values of Uranus (blue dots) and Neptune (orange dots) models used in this study. The bars represent the measured values and their errors for each planet. The hydrostatic $\tilde{J}_4$ (or $\tilde{J}_2$) of the models do not fit for the observed $J^{\rm obs}_4$ (or $J^{\rm obs}_2$) in order to account for the contribution of their dynamical counterparts $J^\prime_4$ (or $J^\prime_2$). Arrows indicate qualitatively the missing dynamical $J^\prime_4$ counterpart to the observed $J^{\rm obs}_4$ according to   Equation \ref{eq:presc}.}
    \label{fig:jj}
\end{figure}

As can be seen from Figure \ref{fig:poly}, the density of the models at the 1-bar pressure level is higher than what is usually inferred for Uranus ($\sim$0.42 kg m$^{-3}$) and Neptune ($\sim$0.45 kg m$^{-3}$) \citep{hueso}.
This is especially true for 3-layered models, as they allow for a larger variation in density in the outer layers.
Interior structure models of Uranus and Neptune based on physical equations of state tend towards a sharper density drop with decreasing pressure in the outer layers. The $\tilde{J}_2$ and $\tilde{J}_4$ values of each model are shown in Figure \ref{fig:jj}. The hydrostatic $\tilde{J}_4$ (or $\tilde{J}_2$) of the models do not fit for the observed $J^{\rm obs}_4$ (or $J^{\rm obs}_2$) by design in order to include the contribution of their dynamical counterparts $J^\prime_4$ (or $J^\prime_2$)

Naturally, the interiors of Uranus and Neptune are not necessarily represented by 3-layer polytropes. This is especially true for the outermost layers, since most interior structure models in the literature assume a more steeply decaying $P-\rho$ scaling than that of polytropes. However, this is of little significance, since in order to probe stronger winds one requires models that have a shallower scaling, as this corresponds to higher hydrostatic $-\tilde{J}_4$ values. 
High hydrostatic surface densities directly imply a planetary mass distribution that is skewed towards the outer polytrope, and therefore yield higher $-\tilde{J}_4$ values. High $-\tilde{J}_4$ values in turn require greater scale heights (i.e. slowly decaying winds) to match the observed $J^{\rm obs}_4$.

\section{Results}
\label{sec:res}
\subsection{Gravitational harmonic perturbations \texorpdfstring{$J_4^\prime$}{Lg}}
\begin{figure*}
    \centering
    \includegraphics[width = 0.91\textwidth]{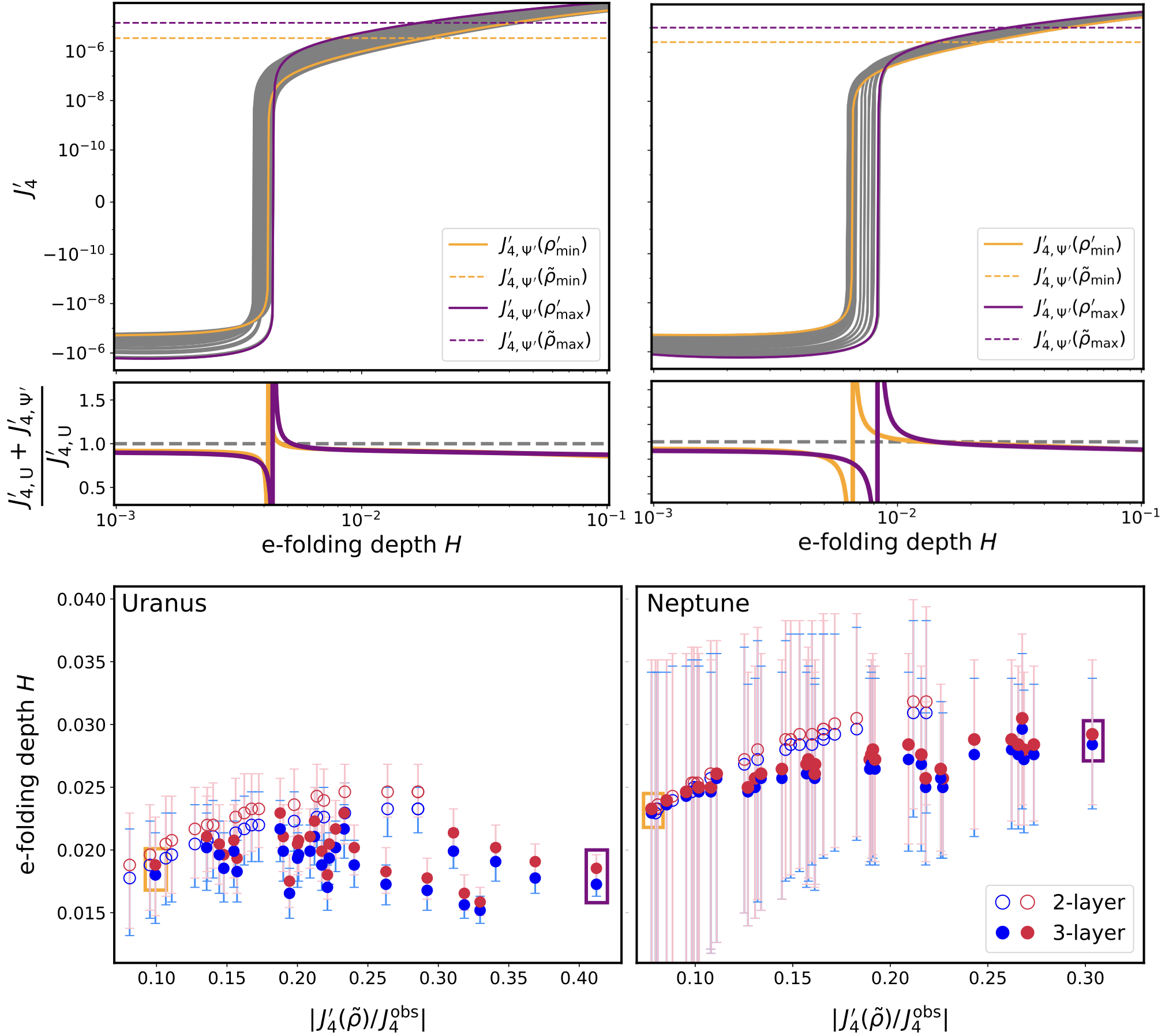}
    \caption{\textbf{Top panels: Contributions to the zonal gravitational harmonic perturbation $J^\prime_4$ for Uranus and Neptune models, as a function of scale height $H$.}  The dashed lines show the difference between the observed and the model 4$^{\rm th}$-degree harmonic: $J^\prime_4 = J^{\rm obs}_4 - \tilde{J}_4$. \textbf{Bottom panels: The relative impact of the self-gravity contribution with respect to the zonal wind contribution as a function of scale height $H$.}  Apart from the singular region where the perturbation changes signs,  self-gravity has a contribution that is about one order of magnitude smaller than that of the zonal winds.  
    \textbf{Bottom panels: The scale height $H$, for which the condition $J^\prime_4(\rho^\prime) = J^\prime_4(\tilde{\rho})$ is satisfied, as a function of the missing dynamical contribution percentage from the model: $|J^\prime_4(\tilde{\rho})/J^{\rm obs}_4| = |1 - \tilde{J}_4/J^{\rm obs}_4|$.}  Since $\tilde{J}_4/J^{\rm obs}_4 > 1$, models lying further out in the x-axis have higher $-\tilde{J}_4$ and require a higher $J^\prime_4(\tilde{\rho})$ to match the observed $J^{\rm obs}_4 = \tilde{J}_4 + J^\prime_4(\tilde{\rho})$. These models correspond to those with lower surface densities, since $r > 0.65 R$ of the planet is represented by a single polytrope.
    Red (blue) dots represent solutions with (without) the contribution from dynamic self-gravity. The uncertainty of the inferred scale height limits for Neptune is larger due to the larger measurement uncertainty in the observed $J^{\rm obs}_4$. Orange (purple) rectangles identify the 3-layer models with the lowest (greatest) $|J^\prime_4(\tilde{\rho})/J^{\rm obs}_4|$ ratio. This color scheme is used in following plots as well.}
    \label{fig:j4}
\end{figure*}

Figure \ref{fig:j4} shows the contributions of both self-gravity and the zonal winds to the perturbation $J^\prime_4$ as a function of the scale height $H$ for the same interior structure model. The effect of self-gravity is typically a factor of ten smaller than that of the zonal winds (except around a small singularity region of $H$, where the zonal wind contribution flips signs). This ratio is in agreement with the findings of \citet{selfg} for the 4$^{\rm th}$-order zonal harmonic.
Figure \ref{fig:j4} shows the scale height of zonal winds as a function of the ratio between the missing dynamical harmonic and the observed total harmonic $|J^\prime_4(\tilde{\rho})/J^{\rm obs}_4|$, for all models. Including self-gravity in the equations increases the maximum scale height by $\sim$10$\%$ compared to the that from zonal winds alone, denoted by red dots. The observational values of the zonal gravity harmonics that we use to account for the missing dynamical contribution $J_4^\prime(\tilde{\rho})$ are given by \citet{j_nep, j_ura}:
\begin{align*}
    J_4^{\rm obs}({\rm Uranus}) &=  (-34.2 \pm 1.3) \times 10^{-6},  \nonumber\\ \nonumber
    J_4^{\rm obs}({\rm Neptune}) &=  (-33.4 \pm 2.9) \times 10^{-6}.
\end{align*}

The scale height for Uranus models are shallower than 0.025$R_{\rm \scriptscriptstyle U}$, and for Neptune 0.03$R_{\rm \scriptscriptstyle N}$. These numbers are in agreement with Ohmic dissipation constraints \citep{soyuer2020}, and the gravity estimates by \citet{kaspi2013}, suggesting that the observed strong zonal winds are likely an atmospheric phenomenon.

\subsection{Effects of bulk rotation period on decay depths}
\citet{helled_shape} suggested that the  rotation periods of Uranus and Neptune  measured by \textit{Voyager II} do not represent the rotation of the deep interiors. By minimizing the dynamical height of zonal winds on Uranus and Neptune, they suggested that Uranus' bulk rotation is $\sim$ $4\%$ faster, and Neptune's bulk rotation is $\sim$ $8\%$ slower than the \textit{Voyager II} period. We repeat our calculations and investigate the sensitivity of the results to the assumed bulk rotation period. 

Since the zonal flows are defined relative to an assumed planetary bulk rotation, any change in the rotation period $T$ would result in an amplitude change in the surface zonal wind profile as:

\begin{equation}
v_{\varphi}^{\text {new}}(\theta)=v_{\varphi}(\theta)+2 \pi\left(\frac{1}{T_{\mathrm{new}}}-\frac{1}{T}\right) R \sin \theta,
\end{equation}
such that the new values imply slower zonal winds for Uranus and faster ones for Neptune.
Using the formulation given in equation (\ref{eq:u}), the addition to the zonal wind profile $\delta U_\varphi$ due to a rotation period change (such that $U_{\varphi, \rm new} = U_\varphi + \delta U_\varphi $) is given by: 
\begin{equation}
    \delta U_\varphi(r,\theta) = (\omega_{\rm new} - \omega) \frac{r^2}{R}\sin\theta \times \mathcal{B}(r,\theta)
\end{equation}
assuming the wind profile penetrates the planet along lines parallel to the rotation axis.

Since this is an additive effect, changes in the rotation period  induce a change  $\delta J^\prime_{4}$ on top of $J^\prime_{4}$ given by: \begin{figure*}
    \centering
    \includegraphics[width = \textwidth]{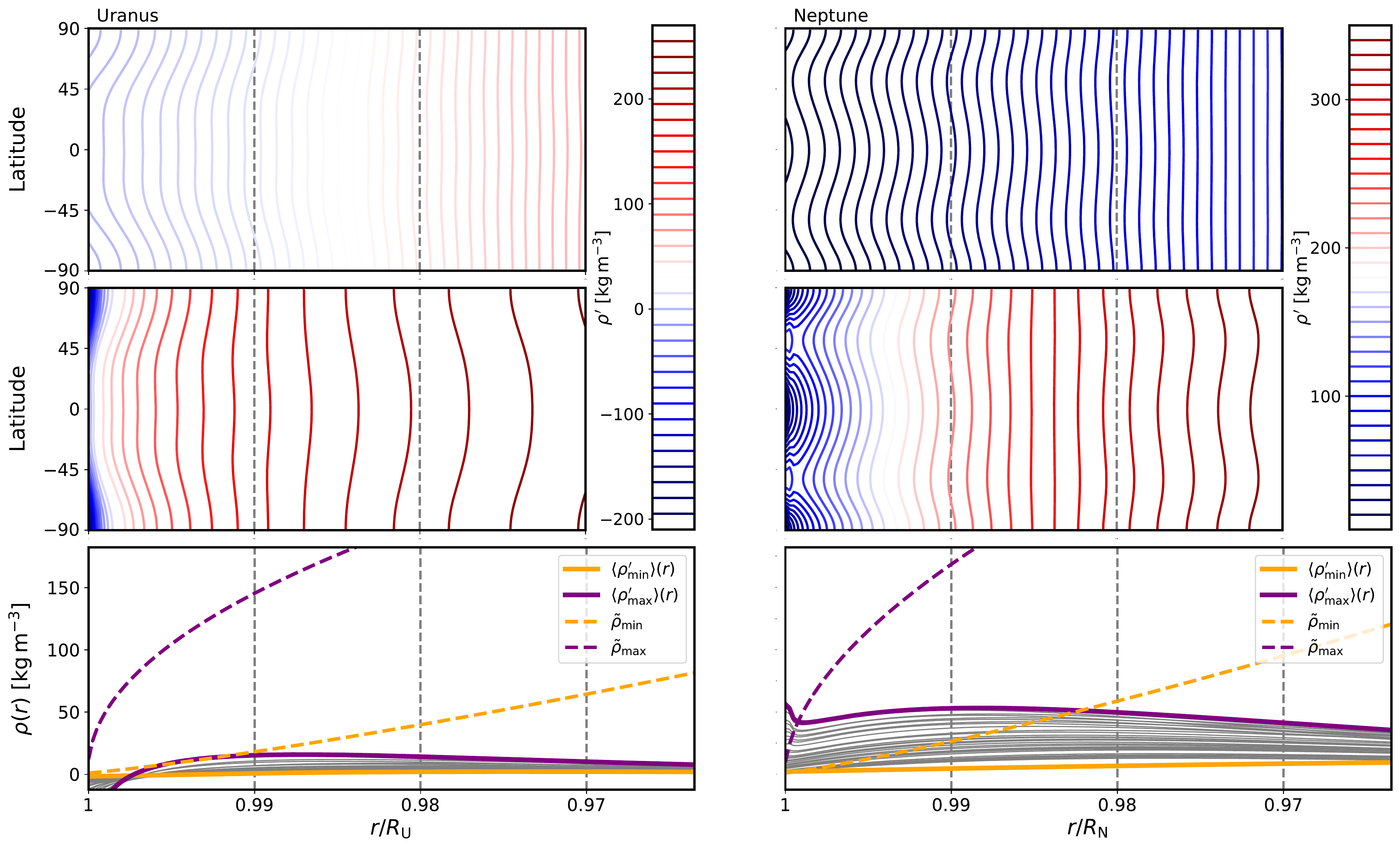}
    \caption{\textbf{Top and center panels: Total density contours as a function of planetary radius $r/R$ and co-latitude $\theta$ of most extreme models of Uranus and Neptune.}  The top (center) panels are the models with the lowest (highest) $\tilde{J}_4$ values (also shown in Figure \ref{fig:j4}). \textbf{Bottom panels: Densities averaged over the co-latitude as a function of planetary radius.} Solid curves show both density perturbations (lowest in orange, highest in purple), as well as all the other models in gray. Dashed curves show the hydrostatic densities of their respective models.}
    \label{fig:rho}
\end{figure*} 
\begin{equation}
    \delta J^\prime_{4} \!\approx\!  \frac{2\Delta \omega}{MR^5}\!\int\limits_V \frac{r^5}{\tilde{g}}  P_4(\cos\theta) \!\int \partial_z (\tilde{\rho}r^2\mathcal{B}(r,\theta)\sin\theta) {\rm d}\theta  {\rm d}V,
\end{equation}
as the inertial effects and self-gravity is of second order in the dynamical density perturbation.
Figure \ref{fig:j4} shows that all models have $H>0.15$, and thus, in the positive end of the $J^\prime_{4}$ (see Figure \ref{fig:j4}). Therefore, it is straightforward to estimate the impact of a change in the planetary rotation period.

For the modified rotation periods presented in \citet{helled_shape}, the  ratio $\delta J^\prime_{4}/J^\prime_{4} < 1$ in Uranus and Neptune for the scale heights probed here. Figure \ref{fig:j4} shows  that doubling the dynamical zonal harmonic $J^\prime_{4}$ has negligible effect on the scale height $H$ due to the  scaling of $J^\prime_{4}$ with $H$. Thus, considering $\delta J^\prime_{4} + J^\prime_{4}$ due to modified rotation periods have same implications for the scale heights of those from models with \textit{Voyager II} measured rotation periods. 
\subsection{Density isosurfaces and planetary shapes}
The density perturbations induced by zonal winds vary with co-latitude and therefore have a direct effect on the shape of the planets.
Figure \ref{fig:rho} shows the  contours of the total density (i.e. hydrostatic density + dynamic density perturbation) of the most extreme models as a function of radius and co-latitude, as well as the individual latitude-averaged density perturbations of all models as function of radius for Uranus and Neptune.

The density isosurfaces  $r(\rho = \rm{const.})$ scale with both the hydrostatic density $\tilde{\rho}$, the decay scale height $H$ and rotation period. However, the shape deviations naturally follow the latitudinal dependence of the wind profile seen in Figure \ref{fig:winds}. 

\begin{figure}
    \centering
    \includegraphics[width = \columnwidth]{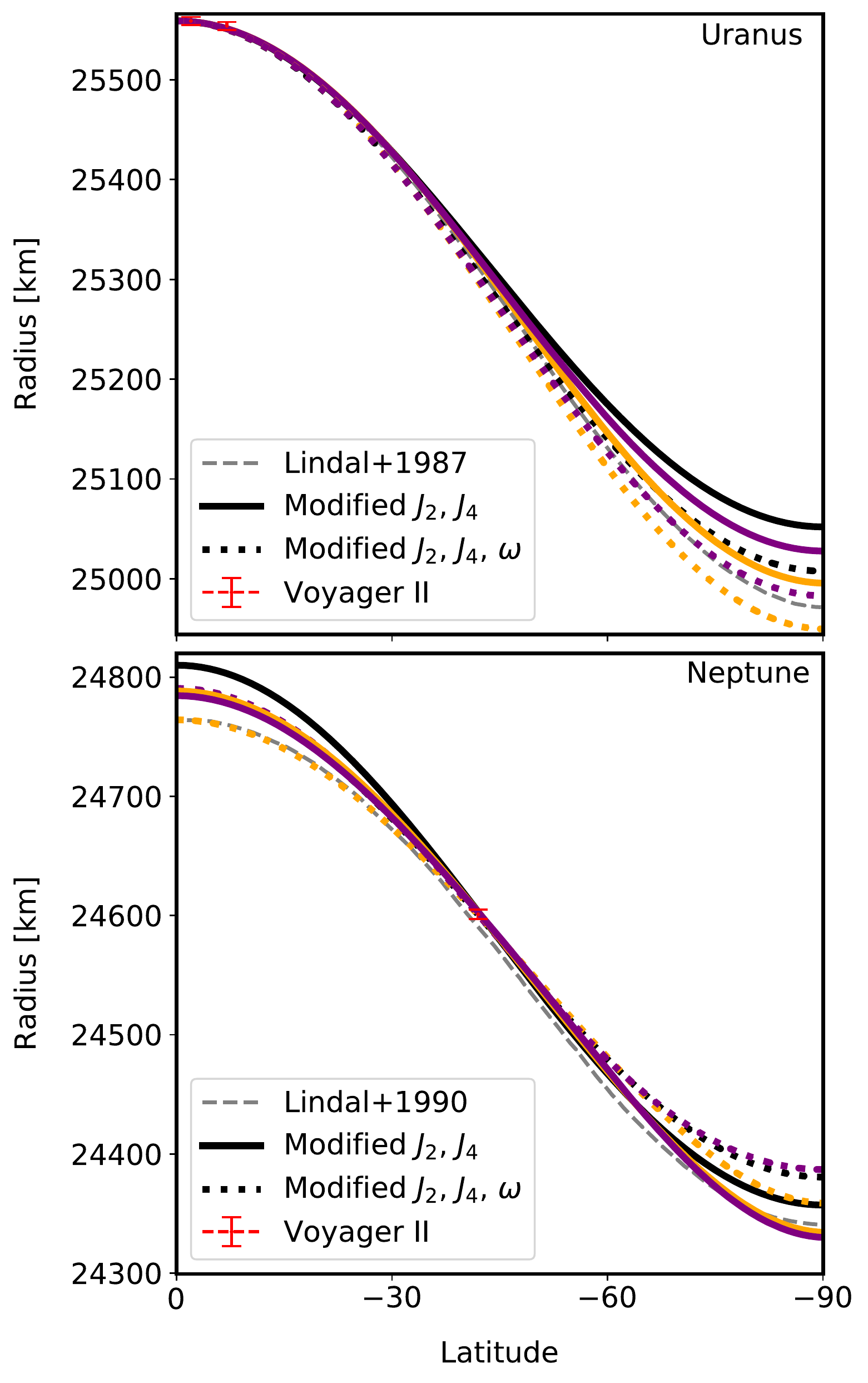}
    \caption{\textbf{The geoid shape as a function of co-latitude for Uranus (top) and Neptune (bottom) for different configurations.} Gray dashed lines show the inferred shapes after \textit{Voyager II} radius measurements \citep{Lindal1987,lindal}. Black solid lines show the same geoid calculation, but with updated $J_2$ and $J_4$ values from \citet{j_ura,j_nep}. Black dotted lines show the geoid shape with updated harmonics as before, but also with an updated rotation period change according to \citet{helled_shape}, where $P_{\rm \scriptscriptstyle U} = 16.58$h and $P_{\rm  \scriptscriptstyle N} = 17.46$h, instead of those in Table \ref{tab:u_tab}. The purple and orange lines correspond to the modifications to the shape given by the black curves due to density perturbations induced by zonal winds, as mentioned in Figure \ref{fig:rho}. Red errorbars show the radius measurements done by \textit{Voyager II}. All curves are normalized such that the \textit{Voyager II} measurements are matched.}
    \label{fig:shape}
\end{figure}

Assuming that these density isosurfaces correspond to the shapes of the planets, we model the 1-bar radius of Uranus and Neptune as a function of co-latitude for different configurations in Figure \ref{fig:shape}. To this end, we calculate the geoid shapes of each planet with 2 different $J_2$ and $J_4$ values; one from the \textit{Voyager II} era \citep{Lindal1987, lindal}, and the updated values that consider ring and satellite dynamics of Uranus and Neptune \citep{ j_nep, j_ura}. Additionally, we consider the rotation period change introduced in \citet{helled_shape}. For all the cases, we take  \textit{Voyager II} radius measurements as a reference and scale the shapes accordingly. Combining the geoid shapes with the wind induced perturbations yield variations in the planetary shapes, especially pronounced in latitudes far away from the measurement positions.

Although the proposed changes in rotation periods in \citet{helled_shape} do not change the maximum scale height of zonal winds in Uranus and Neptune, they would have a measurable effect on the shapes of the planets. Therefore, probing the 1-bar radius at the latitudes where the \textit{Voyager II} radius measurements are extrapolated (i.e. poles of Uranus, poles and equator of Neptune) would yield deeper insights into the dynamics of the planets.
Although our approach for estimating the planetary shapes is a first order approximation, it clearly shows that shape deviations due to zonal winds should be detectable by future missions to Uranus/Neptune. A more rigorous calculation of the shapes would entail generating structure models that consider the zonal wind induced density perturbations in the model generation process, rather than as an additive effect on hydrostatic models. This is an ongoing challenge for interior structure models, and we hope to extend our ToF scheme in the future to account for this non-linearity.

\subsection{Varying \texorpdfstring{$J_2$}{Lg}}
In this section we investigate whether the dynamical correction  to $J_2$ affect our results. In the same vein as before, we generate interior structure profiles of Uranus and Neptune that fit the measured $J^{\rm obs}_4$ zonal harmonic, and then take this as the hydrostatic $\tilde{J}_4$, while keeping $J_2$ as a free parameter. Since generated density profiles are by definition in hydrostatic equilibrium, one obtains $\tilde{J}_2$ as an output of this prescription. We can therefore  attribute the difference between the model output $\tilde{J}_2$ and the observed $J^{\rm obs}_2$ to the missing zonal wind contribution, which we denote as $J^\prime_2(\tilde{\rho})$ (i.e., the dynamical contribution that is missing from the hydrostatic model,
\begin{equation}
J^\prime_2(\tilde{\rho}) = J^{\mathrm{\scriptscriptstyle{obs}}}_2 - \tilde{J}_2.
\end{equation}
We demand that both ways of calculating the perturbation should yield the same results:
\begin{equation}
J^\prime_2(\rho^\prime) \overset{!}{=} 
J^\prime_2(\tilde{\rho}),
\end{equation}
same as in Equation \ref{eq:primee}.

\begin{figure*}
    \centering
    \includegraphics[width = 0.9\textwidth]{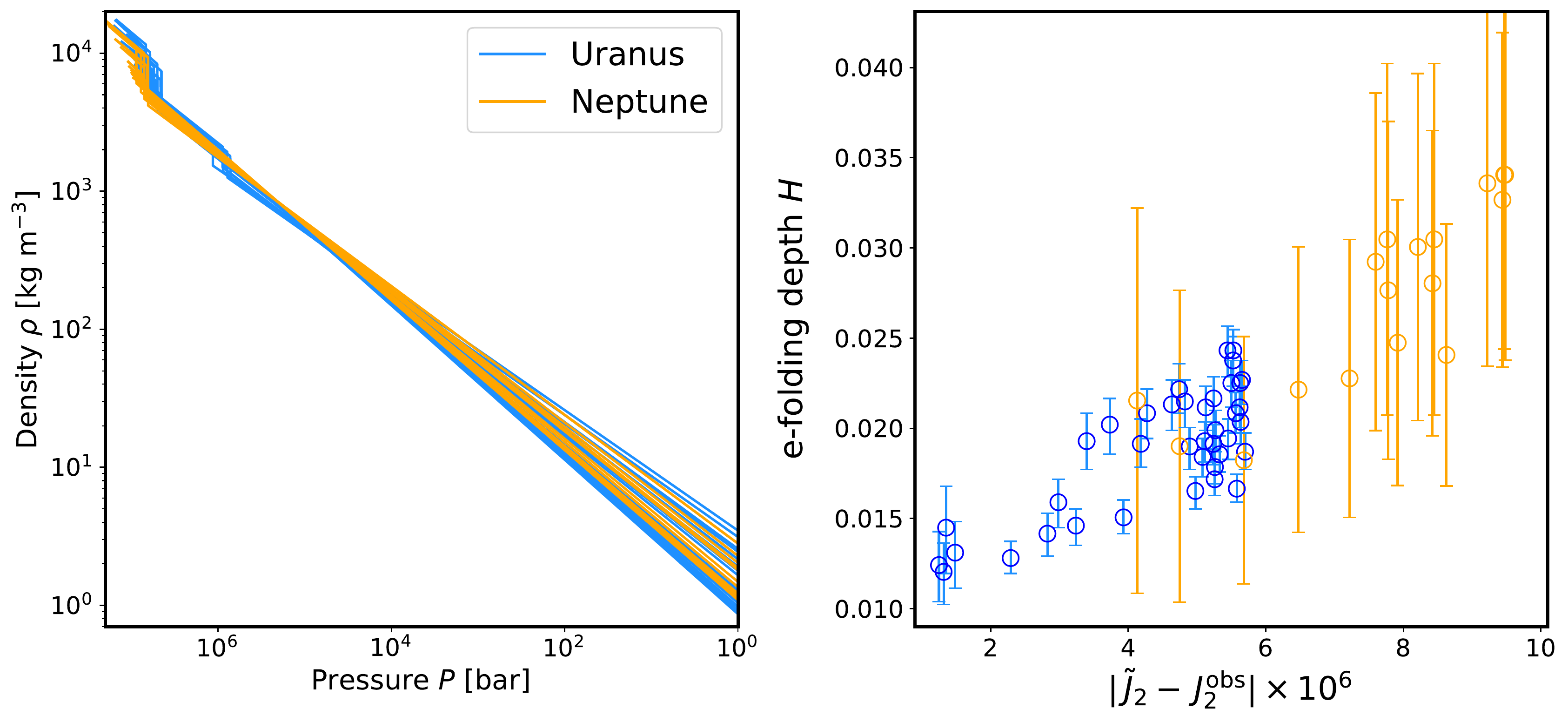}
    \caption{\textbf{Uranus and Neptune interior structure models generated with $\pm 5 \times 10^{-6}$ variation in the observed $J_2$ value (left). Zonal wind decay heights as a function of the $J_2$ offset (right). } Allowing the interior models to have $J_2$ solutions in the $\pm 5 \times 10^{-6}$ range around the mean observed value has the same effect on the zonal wind decay heights as unconstrained, $J_4$.}
    \label{fig:j2last}
\end{figure*}

Figure \ref{fig:j2last} shows  Uranus and Neptune interior structure models generated with $\pm 5 \times 10^{-6}$ variation in the observed $J_2$ value, and the zonal wind decay heights as a function of the $J_2$ offset. We find that allowing the interior models to have $J_2$ solutions in the $\pm 5 \times 10^{-6}$ range around the mean observed value has the same effect on the zonal wind decay heights as not constraining $J_4$. This suggests that the inferred depths of the zonal winds (for the currently available gravity data) are rather robust.

\section{Discussion and Conclusions}
\label{sec:disc}
We presented our analysis on the shallow layer dynamics of Uranus and Neptune caused by  their strong zonal winds.
Our results show that when considering the missing dynamical contribution $J^\prime_4$ to the total observed harmonic $J^{\rm obs}_4$ in our models, the scale height of the zonal winds must be  lower than $\sim$  0.03$R_{\rm \scriptscriptstyle U,N}$ for Uranus and Neptune.  
These values are shallower than what is reported in \citet{kaspi2013}, but are overall in  general good  agreement. Furthermore, we find that the constraints inferred by not fitting $J_4$ in the in models are similar to the one obtained when we  allow  the interior models to have $J_2$ solutions in the $\pm 5 \times 10^{-6}$ range around the mean observed value.

We also show that the proposed bulk rotation period changes to Uranus and Neptune have little to no effect on the maximum wind scale heights, but do have a measurable effect on the shapes of the planets. This effect, combined with the shape deviations due to zonal winds should be detectable by future Uranus and Neptune orbiters. 

The contribution of self-gravity to the total dynamical zonal harmonic is about one order of magnitude smaller than that of the zonal winds. This results is also in agreement with the expected magnitude of self-gravity on the dynamical harmonic $J^\prime_4$ \citep{selfg}. Although, self-gravity has little to no effect on inferring the maximum decay scale heights, it should still play a  role in interpreting future measurements of the gravitational fields of Uranus and Neptune.

The intensity of the zonal winds in Uranus and Neptune have direct implications on the strength of the secondary magnetic fields due to the $\omega$-effect induced magnetic field generation via zonal winds \citep{soyuer2021}. In gas giants, \citet{hao} show  that  these magnetic field perturbations, spatially correlated with zonal winds should be detectable. A future mission to Uranus/Neptune  could be able to probe this phenomenon.

Uranus and Neptune are still under-explored compared to the gas giants, and the scientific yield of prospective \textit{in-situ} missions to these planets are being rigorously examined by the planetary science community \citep[see e.g.][]{exp_hof,exp_helled, exp_dahl, exp_fle, fletcher,  whitepaper, whitepaper2, GW, zwick}. Furthermore,  various mission concepts have already been suggested \citep[e.g.][]{quest, inpro, whitepaper3, simon, odys}.
Currently we do not have measurements of the higher order even harmonics ($J_{6,8,10...}$) and odd harmonics ($J_{2i+1}$) for  Uranus and Neptune.
We strongly support future measurements of  higher order gravitational harmonics by future missions that would improve our understanding of the  atmosphere dynamics and  interior structures of  Uranus and Neptune. 

\begin{acknowledgments}
    DS thanks Hugues de Laroussilhe for fruitful discussions. DS is grateful for the moral support by Ekin Ar{\i}n. We acknowledge support from the Swiss National Science Foundation (SNSF) under grant \texttt{\detokenize{200020_188460}}. 
\end{acknowledgments}
\newpage

\bibliography{sample631}{}
\bibliographystyle{aasjournal}

\appendix
\section{Derivation of \texorpdfstring{$\Psi^\prime$}{Lg}}
\label{sec:app}
The Poisson equation $\nabla^2 \Psi(\bm{r}) = 4\pi G \rho(\bm{r})$ has the classical solution given in many textbooks:
\begin{equation}
\Psi(\bm{r})=4 \pi G \int  \Gamma(\bm{r}, {\tilde{\bm{r}}}) \rho(\tilde{\bm{r}}) {\rm d}V,
\end{equation}
where $\Gamma(\bm{r},\tilde{\bm{r}}) = -1/(4 \pi |\bm{r} - \tilde{\bm{r}}|)$ is the Green's function for the Laplace operator, which can be expanded in Legendre polynomials $P_l(\cos\theta)$, such that
\begin{equation}
\Gamma(\bm{r}, \tilde{\bm{r}})=-\frac{1}{4 \pi} \sum_{l=0}^{\infty} \frac{r_{<}^{l}}{r_{>}^{l+1}} P_{l}(\cos\theta) P_{l}(\cos\tilde{\theta}).
\label{eq:gamma}
\end{equation}
Here, $r_{>}$ and $r_{<}$ represent the larger and the smaller of the radii $r$ and $\tilde{r}$, respectively. The extension of this method to the inhomogeneous Helmholtz equation follows from the same principle that certain radial functions times Legendre polynomials $f(r) P_l(\cos\theta)$ are the eigenfunctions of the Laplace operator. For the Poisson equation, the radial functions $f(r) = r^l$ and $f(r) = r^{-(l+1)}$ satisfy this condition, hence the appearance of the radial terms in equation (\ref{eq:gamma}).

For the inhomogeneous Helmholtz equation that we have in equation (\ref{eq:helmholtz}), one needs a different set of radial orthonormal functions. Spherical Bessel functions of the first kind $j_l$ satisfy: 
\begin{equation}
\nabla^{2} j_{l}(k_{l n} r) P_{l}(\cos\theta)=-k_{ln}^2 j_{l}(k_{l n} r) P_{l}(\cos\theta),
\end{equation}
as explained in \citet{selfg}. Here $k_{ln}$ are chosen so that $j_l(k_{ln}R)$ satisfy the boundary conditions of the Poisson equation. Using the recurrence relations of spherical Bessel functions, \citet{selfg} expresses this boundary condition as: 
\begin{equation}
    j_{l-1}(k_{ln}R) = 0.
\end{equation}
Lastly, with the normalization constants
\begin{equation}
N_{l n}=\sqrt{\frac{2}{R^{3} j_{l}^{2}(k_{l n} R)}},
\end{equation}
the potential $\Psi^\prime$ can be expressed as a series of normalized spherical Bessel functions $j^\star_{ln}(r) = N_{ln} j(k_{ln}r)$:
\begin{equation}
\Psi^{\prime}(\bm{r})=\sum_{n=1}^{\infty} \sum_{l=1}^{\infty} \Psi_{l n}^{\prime} j_{l n}^{\star}(r) P_{l}(\cos\theta).
\end{equation}
The coefficients $\Psi^\prime_{ln}$ are then found via the spectral decomposition, acquired by plugging the potential into the inhomogeneous Helmholtz equation (\ref{eq:helmholtz}), and exploiting the orthonormality of $j^\star_{ln}(r)P_l(\cos\theta)$:
\begin{equation}
\Psi_{l n}^{\prime}=-\frac{G(2 l+1)}{k_{l n}^{2}-\mu} \int \rho^\prime_U(\bm{r}) j_{l n}^{\star}(r) P_{l}(\cos\theta) {\rm d}V,
\end{equation}
which only depends on the choice of  $\mu$ and the density perturbation due to zonal winds $\rho_U$. Varying the zonal wind profile $\rho_U$ has a marginal effect on the the outcome of the self-gravity term, especially for low harmonic degrees \citep{selfg}. Thus, the exact choice of $\mathcal{U}(U_\varphi, \tilde{\rho})$ (and thus $\rho^\prime_U$) is not crucial for evaluating the self-gravity contribution to the density perturbations $\rho^\prime_{\Psi^\prime}$, as long as reasonable profiles are adopted. As demonstrated by \citet{selfg}, and as we show below, the contribution of self-gravity to the zonal gravitational harmonic perturbations $J_{4,\Psi^\prime}$ is generally one order of magnitude smaller than that of the zonal winds.



\end{document}